\documentclass[12pt,preprint]{aastex}
\usepackage{natbib}
\usepackage{amsmath,amsfonts}
\usepackage{color}

\def\bE{{\bf E}}

\def\bu{{\bf u}}

\def\br{{\bf r}}
\def\br'{{\bf r'}}

\def\N{\mathbb{N}}

\def\Na{{\boldsymbol{\nabla}}}
\def\Nas{{\boldsymbol{\nabla}_{s}}}

\def\div{{\boldsymbol{\nabla}}\cdot}

\def\rot{{\boldsymbol{\nabla}}\times}

\newcommand{\abs}[1]{|#1|}

\def\xu{{\bf{\hat x}}}
\def\yu{{\bf{\hat y}}}
\def\zu{{\bf{\hat z}}}

\newcommand{\dep}[1]{\partial_{#1}}

\def\Nu{{\hat{\bf N}}}

\def\B{{\bf B}}

\def\Bz{B_{z}}

\def\bBpo{{\bf B}_{\pi}}

\def\bBop{{\bf B}_{\sigma}}

\def\Wpo{W_{\pi}}
\def\Wop{W_{\sigma}}

\def\po{\pi}

\def\E{{\bf E}}

\def\v{{\bf v}}

\def\N{{\bf N}}

\def\B{{\bf B}}

\def\v{{\bf v}}

\def\B{ {\bf B} }

\slugcomment{To be published in ApJ Letters}

\shorttitle{Coronal Mass Ejections Initiation}
\shortauthors{Amari et al.}

\begin{document}

\title{CORONAL MASS EJECTION  INITIATION:  \\
 On the nature of the Flux Cancellation Model }

\author{T. Amari}
\affil{CNRS, Centre de Physique Th\'eorique de l'Ecole
Polytechnique, F-91128 Palaiseau Cedex, France}
\email{amari@cpht.polytechnique.fr}
\author{J.-J. Aly}
\affil{AIM - Unit\'e Mixte de Recherche CEA - CNRS - Universit\'e Paris VII 
- UMR n$^{\tiny{0}}$ 7158, Centre d'Etudes de Saclay, F-91191  Gif sur Yvette Cedex, France}
\author{Z. Mikic and J. Linker}
\affil{Predictive Science Inc., San Diego, CA 92121, USA}

\begin{abstract}
We consider a three-dimensional bipolar force-free magnetic field with non zero magnetic helicity,
occupying a half-space, and study the problem of its evolution driven by an imposed photospheric flux decrease.
For this specific setting of the Flux Cancellation Model describing coronal mass ejections occuring in active regions,  
we address the issues of the physical meaning of flux decrease, of the influence on field evolution of the size of the domain over which this decrease is imposed, and of the existence of an energetic criterion characterizing the possible onset of disruption of the configuration. We show that: (1) The imposed flux disappearance can be interpreted in terms of transport of positive and negative fluxes towards the inversion line, where they get annihilated.  (2) For the particular case actually computed, in which the initial state is quite sheared, the formation of a twisted flux rope and the subsequent global disruption of the configuration are obtained when the flux has decreased by only a modest amount over a limited part of the whole active region. (3) The disruption is produced when the magnetic energy becomes of the order of the decreasing energy of a semi-open field, and then before reaching the energy of the associated fully open field. This suggests that the mechanism leading to the disruption is nonequilibrium as in the case where flux is imposed to decrease over the whole region.
\end{abstract}


\section{INTRODUCTION}

Magnetic flux cancellation (FC) is a well documented solar photospheric phenomenon \citep{Welsch06} which plays the key role in the Flux Cancellation Model (FCM). FCM has been introduced initially to explain flux rope and prominence formation \citep{VanBallegooijenMa89}, and applied later on for modelling large scale eruptive events such as Coronal Mass Ejections (CMEs) occurring in active regions (see, e.g., \citet{ForbesLiChEtal06} for a review). In the latter context, it has been actually developed in two settings differing from each other essentially by the way FC is introduced: FC is either taken to be a physical consequence of the photospheric turbulent diffusion, or just imposed as a time dependent boundary condition describing a decrease of the flux according to some prescribed law. 
In the former case, first considered in \citet{AmariLuAlMiLi03b} and later on in \citet{YeatesMa09}, \citet{YeatesAtNaEtal10}, and \citet{AuToDemDel09}, the dispersive effect of the turbulence on the flux of an active region leads to the bringing together on the inversion line of some amounts of flux of both polarities, which thus get annihilated by small scale mixing. This type of FCM is of course particularly relevant when one wants to explain the CMEs which occur during the decaying phase of an active region. The second type of FC implementation was first applied to study the possible disruption of 2D bipolar configurations \citep{ForbesPriest95}, and later on of 3D bipolar configurations  \citep{AmariLuMiLi00,LinkerLiMiAm01} and 3D quadrupolar ones \citep{AmariAlMiLi07}.
\par

Although previous works have led to a good understanding of FCM in the ``flux decreasing setting'', there are many points which are still unclear.
(1) The disappearance of flux on the boundary is imposed as a mere mathematical boundary condition, and 
doubts have been sometimes expressed about the possibility that it be related to some physical process. To keep viable this setting of the FCM, we thus need to provide a physical interpretation of flux decrease compatible with the observations. For instance, the latter show FC to be often associated to the mutual annihilation of opposite polarity flux elements brought into contact by photospheric motions, and an interpretation in terms of flux transport and annihilation on an inversion line would be certainly adequate. An other possibility would be to interpret FC in terms of emergence through the photosphere of either an U-loop or a bipolar loop, flux decrease occuring in the latter case once the magnetic axis has started emerging
(see \citet{Fan01a} and \citet{AmariLuAl05} for simulations, and \citet{LopezFuentesDeManVan00} for observations). 
(2) Previously, flux decrease has been imposed to occur over the whole active region, or at least  the whole central bipolar part in the case of a quadrupolar region \citep{AmariAlMiLi07}. In many cases, however, one may expect FC to occur over only a small part of the region, and the question arises of the possibility of still triggering a disruption in that case. 
(3) If a disruption actually occurs, we have finally to address the question of the nature of its trigger. Basically, there are three possible mechanisms  \citep{AmariAl09}: Non Equilibrium (NE), Quasi Non Equilibrium (QNE), meaning equilibrium too far to be accessible, and Unstable Equilibrium (UE). For instance, we found QNE to be responsible for the transition to very fast expansion exhibited by a flux rope twisted by boundary motions  \citep{AmariLuAlTa96b}. In the case of a bipolar  \citep{AmariLuMiLi00} or quadrupolar \citep{AmariAlMiLi07} configuration submitted to FC, on the contrary, we found NE to be at the origin of the eruption. This conclusion was established by showing that the magnetic energy of the system exceeds the energy of the associated totally open field in the former case, and the energy of a partially open field (having its open lines being connected only to the part of the boundary where strong currents once developed) in the latter one.  
Then we should look in particular for the existence of a similar energy criterion if a disruption occurs as a consequence of partial FC. 
\par
 
The aim of the present Letter is to address these important issues in the case where the initial configuration is force-free and has a nonzero magnetic helicity. Physically, such a field may be thought of either resulting from the emergence of a twisted flux rope (TFR) through the photosphere, or being the remnant of a configuration which has previously generated an eruptive event without fully relaxing to a potential field \citep{AmariLu99}, or to be produced by photospheric twisting motions. It is worth insisting on the fact that the evolutions we consider are only driven by FC, in opposition to some other works \citep{vBallegooijenPrMa00,AuToDemDel09} in which  FC is applied along with shearing motions.

\section{MODEL AND INITIAL CONFIGURATION}

In our numerical model, the ``coronal half-space'' $\{z>0\}$ above an active region is represented by a large computational box 
$\Omega_{h}=[-20,20]\times[-20,20]\times[0,40]$, equipped with a non uniform mesh of 141x131x121 nodes. 
The used physical quantities are dimensionless. In particular the Alfven 
crossing time $\tau_{A}$ is taken as the time unit. $\Omega_{h}$ contains a perfectly conducting low density plasma with an embedded magnetic field $\B$. For $t\geq 0$, this system is brought into a two-stages MHD evolution: a preliminary stage, not claimed to represent an actual coronal evolution, which produces
a force-free equilibrium with a nonzero helicity, and a second stage, which is the physically relevant one, in which the previous field is taken as the initial state of a FC driven evolution.
Both phases are controlled by imposing on the ``photospheric'' boundary 
$S_{h}$ of $\Omega_{h}$ (on which $z=0$) the tangential component of the electric  field, $\E_{s}$, a procedure which is known from our previous works to lead to a well posed problem. Quite generally, $\E_{s}$ can be Helmholtz decomposed according to
\begin{eqnarray}
	c\E_{s} =  \Nas\phi + \Nas\psi\times\zu ,
	\label{HDecomp}
\end{eqnarray}
where $\Nas=\xu\,\dep{x}+\yu\,\dep{y}$, and $\phi(x,y,t)$ and $\psi(x,y,t)$ are solutions, respectively, of the equations
$\Nas^{2}\phi = c\Nas\cdot\E_{s}$ and 
$-\Nas^{2}\psi = c\zu\cdot\Nas\times\E_{s}$. Using  Faraday law thus leads to
\begin{eqnarray}
	\dep{t}\Bz = -c\div(\E_{s}\times\zu) = \Nas^{2}\psi .
	\label{Induction}
\end{eqnarray}
The  MHD equations are solved by our numerical algorithm \citep{AmariLuJo99, AmariLuAlTa96b}. 
Small values are choosen for the dissipation coefficients:
$\nu=10^{-2}-10^{-3}$ for the kinematic viscosity, and $\eta$ not larger than $10^{-4}$ for
the resistivity, giving for our mesh resolution a Lundquist number not smaller than  $10^{4}$.
\par

\begin{figure}
\epsscale{0.95}
\plotone{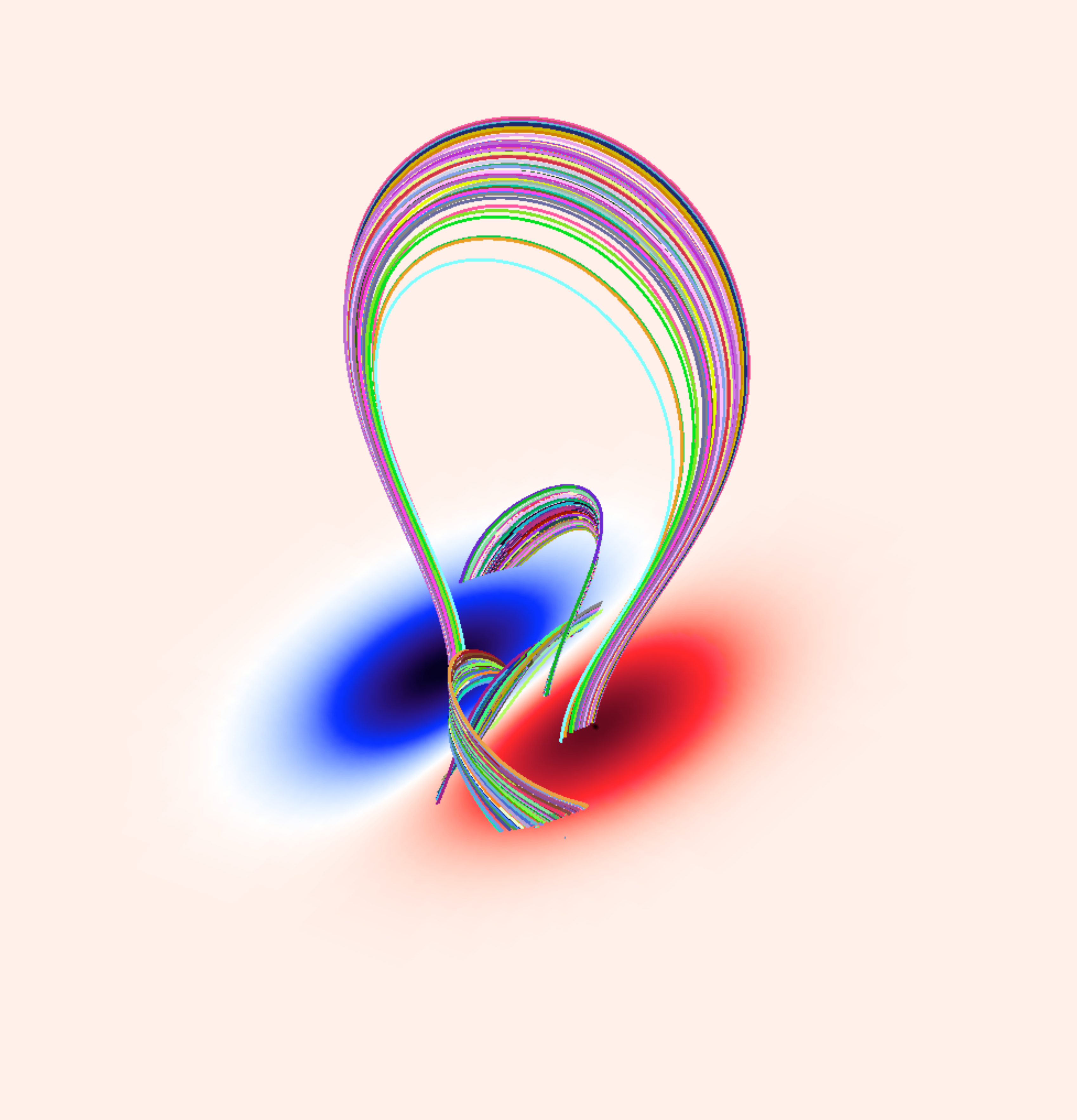}
\caption{Selected field lines of the initial force-free configuration
reached after a shearing-twisting
phase followed by a viscous relaxation. Strong shear is accumulated along the neutral line.
\label{fig1} }
\end{figure}

In the first stage, we start at $t=0$ from a bipolar potential magnetic field $\bBpo = \Na V_{\po}$. $V_{\po}$ is the solution of a Dirichlet-Neuman boundary value problem for Laplace equation, in which we impose in particular 
the condition
\begin{eqnarray}
	B_{\po z}(x,y,0) = q(x,y) 
	= e^{-x^{2}/\sigma_{x}^{2}} \left[e^{-(y-y_{c})^{2}/\sigma_{y}^{2}} -e^{-(y+y_{c})^{2}/\sigma_{y}^{2}}\right]
	\label{}
\end{eqnarray}
on $S_{h}$,
with $y_{c}=-0.8, \sigma_{x}=1, \sigma_{y}=2$.
For $0\leq t\leq 400$, we apply on $S_{h}$, as in \cite{AmariLuAlMiLi03a}, 
the tangential electric field $\E_{s}$ associated to $\psi=0$ and a $\phi$ such that 
$\phi'(B_{z})=B_{z}\Phi'(B_{z})$, with $\Phi(B_{z})$ a prescribed function. This generates slow shearing/twisting 
$B_{z}$-preserving
ideal motions of the magnetic footpoints at the velocity $\v_{s}=c\Nas\phi\times\zu/\Bz$
(with $\max(v_{s})=10^{-2}$). 
A neighbouring nonlinear
force-free equilibrium is next reached by performing for $400\leq t\leq 800=t_{0}$ a viscous relaxation phase during which 
$\phi=\psi=0$ and $\Bz$ is still conserved on $S_{h}$.
As shown on Figure \ref{fig1}, this equilibrium is sheared along the neutral line, and because of twist, it exhibits away from that line the presence of strong electric currents correlated with the typical sigmoidal structure. 
It is worth noticing that, for this equilibrium,  
we have $supp(FFF)\subset supp(\Bz)$, where $supp(FFF)$ and $supp(\Bz)$ denote the regions where, respectively, electric currents and $\Bz$ are strong enough.

\notetoeditor{Figure\ref{fig1} should appear here}

\section{PARTIAL FLUX CANCELLATION}

For driving the FC phase, we fix $\E_{s}$ 
on $S_{h}$ by setting $\phi(x,y,t) = 0 \,$ and by taking $\psi$ to satisfy 
$\Nas^{2}\psi(x,y,t) = -\mu\zeta(y)\Bz(x,y,0,t_0)=-\mu\zeta(y)q(x,y)$ on $S_{h}$, with $\mu=10^{-2}>0$, and 
$\zeta(y)=[1 + \tanh[(y_1-y)/d] ] [1 + \tanh[(y-y_0)/d] ]/4$. 
$\zeta$ exhibits a "plateau" of height 1 in the interval $[y_0, y_1]$ and falls down to zero in a layer of thickness $d$. It is used
as a mask controlling the size of the cancellation support, 
$supp(FC)$. We use here the particular values $y_0=-0.5, y_1=0.5, d=0.1$, whence $supp(FC)\subset supp(FFF)\subset supp(\Bz)$ as the width of each Gaussian polarity is approximatly equal to $2$ and decreases at a slower rate.
The case considered in \citet{AmariLuMiLi00} (where FC is enforced over the whole bipolar region)
corresponds to $\zeta=1$. Using 
Eq. (\ref{Induction}), we have $\dep{t}\Bz(x,y,0,t)=-\mu\zeta(y)q(x,y)$, whence after an immediate integration
\begin{eqnarray}
	B_{z}(x,y,0,t) = q(x,y)[1-\mu\zeta(y)(t-t_{0})] .
	\label{Bz(x,y,0,t)}
\end{eqnarray}
Then $\Bz$ suffers a linear decrease on $S_{h}$. But this does not lead to any specific problem as the reduction factor 
$[1-\mu\zeta(y)(t-t_{0})]\geq 0.62 > 0$, 
the FC phase being limited to a duration of about $38$ (see below).
All along that phase, we regularly apply viscous relaxation runs to check if a neighbouring equilibrium exists or not. We found this strategy to be more appropriate than the one used in  \cite{AmariAlMiLi07}, where the need to effect episodic relaxations was bypassed by choosing a much smaller value of 
$\mu$ ($\mu=10^{-4}$). 
\par

As stated in Section 1, it is crucial for the validity of the model that the mathematically imposed flux decrease be interpretable in physical terms. Let us show that it can be actually considered to result from the transport of opposite polarities fluxes towards the inversion line $I$ where they annihilate indeed. In any case, we can introduce on $S_{h}$ an horizontal 
velocity of magnetic flux transport, ${\bu}$, by setting  
$c\E_{s} + B_{z}\bu\times\zu = 0$. Using the Helmholtz decomposition (\ref{HDecomp}) of $c\E_{s}$ and 
Eq. (\ref{Bz(x,y,0,t)}), we obtain at once
\begin{eqnarray}
	\bu(x,y,t) = -\Nas \psi (x,y) / [q(x,y) (1 - \mu\zeta(y)(t-t_{0}))] .
\end{eqnarray}
During the FC phase, $\bu$ appears to be just continuously rescaled by a time dependent factor on most of $supp(FC)$, 
where $ \zeta = 1$, and to remain invariant outside $supp(FC)$, where $\zeta=0$. 
Interestingly, we see that the cancellation flow support, 
$supp(CF)$, coincides approximately with $supp(B_{z})$, and is then not reduced to
$supp(FC)$ for a localized FC. The velocity $\bu$ computed at some time 
is shown on Figure \ref{fig2ab}a, and it is clearly seen to be directed indeed towards $I$ in a neighbourhood of that line. Note that we have removed from the plot the parts very close to $I$ and very far from the spots, respectively, where $u$ becomes very large. The flux $B_{z} \bu$ of $\Bz$ keeps however a finite value on $I$, on which $\Nas\psi$ does not vanish, and the flow thus continuously advects magnetic flux onto that line, where it disappears. 
On the other hand, it should be noted that the velocity pattern at any other time during the FC phase is identical to the one shown on Figure \ref{fig2ab}a because of the simple scaling property proved above.

\notetoeditor{Figure\ref{fig2ab} should appear here}

That flux decrease may be associated with flux annihilation on the inversion line $I$ can be also simply proved in a quite general way as follows. Let us consider for a little while an arbitrary initial distribution $\Bz(t_{0})$ on $S_{h}$, and  prescribe a flux decrease by setting $\partial_t B_{z} = f_{c}(x,y,t)$, with $f_{c}\geq 0$ on $S_{h}^{+}$, where $\Bz>0$ 
($f_{c}\leq 0$ on $S_{h}^{-}$, where $\Bz<0$),
and  $\int_{S_{h}} f_{c} ds = 0$. We define as above the flux transport velocity $\bu$ by  
$B_{z}\bu = c\E_{s}\times\zu=\Nas\phi\times\zu - \Nas\psi$. Then we have 
\begin{eqnarray}
	\int_{I}\Bz u_{N} dl = \int_I (\dep{l}\phi-\dep{N}\psi)dl = - \int_{S_{h}^{+}} f_{c}ds > 0,
	\label{}
\end{eqnarray}
where $\N$ is the exterior normal on  $S_{h}$ to $\partial S_{h}^{+}$ (and then to $I$), and $u_{N}=\bu\cdot\N$. For estimating the second member, we have used  Gauss theorem, the equation $\Nas^{2}\psi=f_{c}$ (which results from Eq. (\ref{Induction})), and 
assumed $\dep{N}\psi$ 
and $\phi$ to vanish on the far part of the boundary. We can thus conclude that,  "in the mean sense ", we have in a neighbourhood of $I$ (into which $\N$ has been extended)
$u_{N} > 0$ on the $+$ side and $u_{N} < 0$ on the $-$ side, which corresponds indeed to a finite amount of fluxes of both signs being transported towards $I$, where they cancel. But of course, for arbitrary functions $\Bz(t_{0})$ and 
$f_{c}$, and associated sinuous shape of $I$, we expect to have in general $u_{N}<0$ ($u_{N}>0$) on the $+$ ($-$) side of some ``little'' part of $I$, i.e.,  
local emergence of flux, and FC holds only globally. 
That this does not happen in the particular case considered in this Letter is due to the fact that we consider a configuration with antisymmetric flux distribution ($q(x,-y)=-q(x,y)$, whence $I=\{y=0\}$), and 
impose $f_{c}= -\mu\zeta q$, with $\zeta$ symmetric ($\zeta(-y)=\zeta(y)$). 
Finally, we note the following relation, which holds for the latter specific choice of $f_{c}$. Consider an arbitrary curve $C$ surrounding the positive part of $supp(FC)$, with external normal $\Nu$. Setting $\Phi(t) = \int_{supp(FC)^{+}}B_{z} ds$ and proceeding as above, we get 
\begin{eqnarray}
	\int_{C}B_{z}u_{N} dl = -d_t\Phi = \mu\int_{supp(FC)^{+}}\zeta q ds  \simeq \mu\Phi(t_{0})
	\label{}
\end{eqnarray}
Increasing the size of $supp(FC)$ increases $\Phi(t_{0})$ and thus the mean velocity $<u_{N}>$.

For a turbulent diffusion driven evolution \citep{AmariLuAlMiLi03b}, the arguments above, based on our Helmholtz decomposition of $\bE_{s}$, also apply, thus providing a formal proof that cancellation flows are associated to that process too.

\notetoeditor{Figure\ref{fig2ab} should appear here}

\section{TWISTED FLUX ROPE, DISRUPTION, AND TRIGGERING MECHANISM}
\label{results}

We now describe the evolution of the coronal field driven by FC.
\par

(1) During a first phase, there is a transition from an arcade topology to a twisted flux rope (TFR) topology. This transition is not instantaneous. Rather we observe more and more field lines close to $I$ forming dips and a TFR progressively grows from those as flux is advected  
towards $I$. A coherent TFR exists at $t=t_{FR}$. In particular J-shape loops have "merged", generating the inverse topology characteristic of the TFR, with dips able to support prominence material, as seen on Fig. \ref{fig3}. Such a feature was also found in the simulations reported in \citet{AmariLuMiLi00}. During the transition, the mutual helicity of the two J-shape loops is converted into the TFR self-helicity. 

\notetoeditor{Figure\ref{fig3} should appear here}
 
\begin{figure}
--\epsscale{0.9}
\plotone{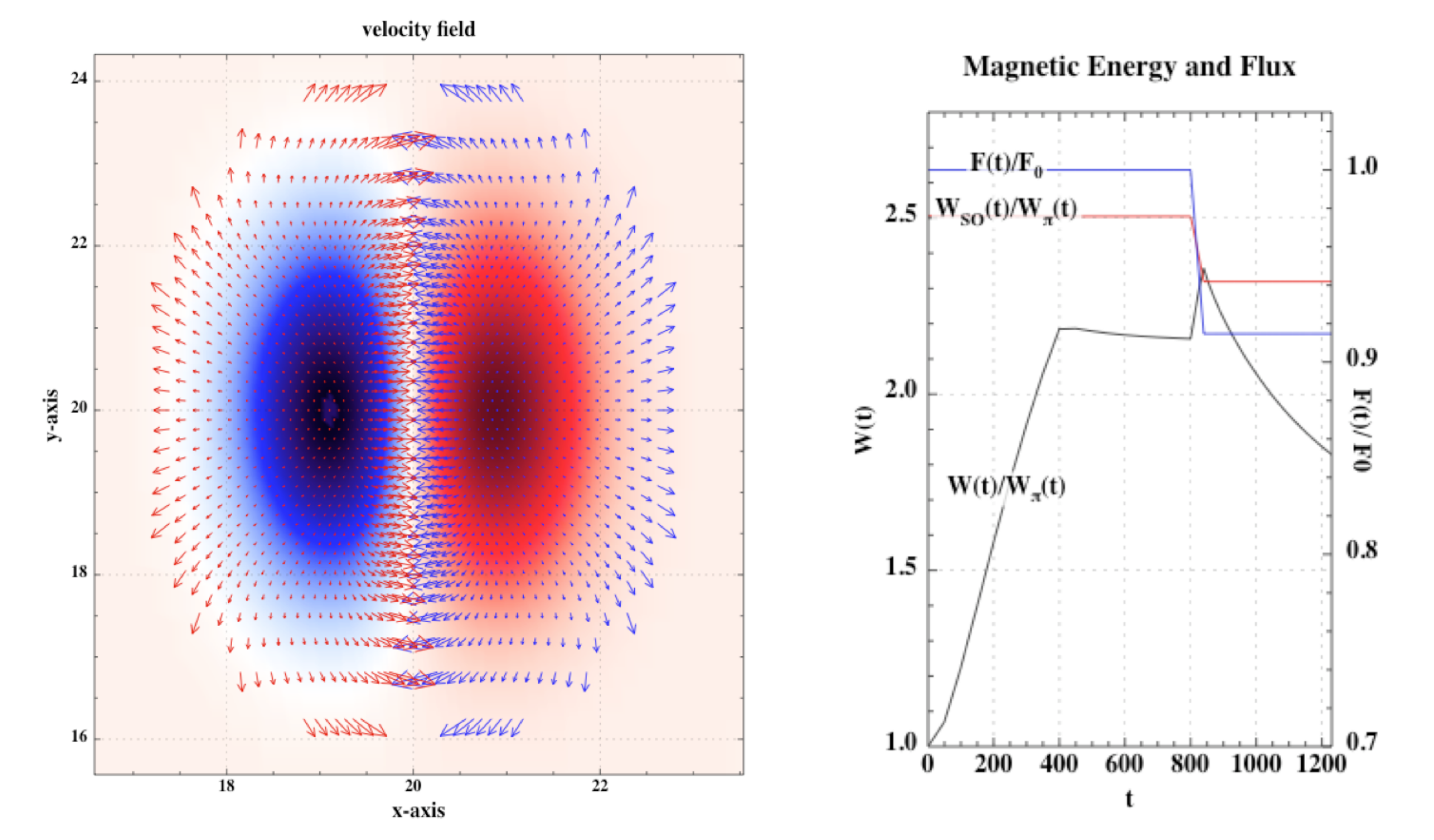}
\caption{(a) Plot of the velocity field associated with the imposed flux decrease. This flow is indeed a cancellation flow as it brings flux elements of opposite polarities into contact along the inversion line, where they get annihilated.
(b) Time variations of some important quantities during the 
phase of twisting by boundary motions of the initial potential
field ($0\leq t\leq400$), the phase of viscous relaxation
($400\leq t\leq 800$), the phase of FC ($800\leq t\leq 838$), and the nonequilibrium phase ($838\leq t$). Are represented the normalised unsigned flux 
$F(t)/F_{0}$ through $S_h$, 
the free magnetic energy measure $W(t)/W_{\pi}(t)$ (which increases during the FC phase), and the free
energy measure $W_{SO}(t)/W_{\pi}(t)$ (which decreases during FC phase)  of a semi-open field which has all its open lines originating from the region 
on which the shear/twist is distributed. The last two curves intersect at some critical time, corresponding to a FC of less than 
$6\%$, beyond which the global disruption occurs.
\label{fig2ab}}
\end{figure}

\begin{figure}
\epsscale{1.}
\plotone{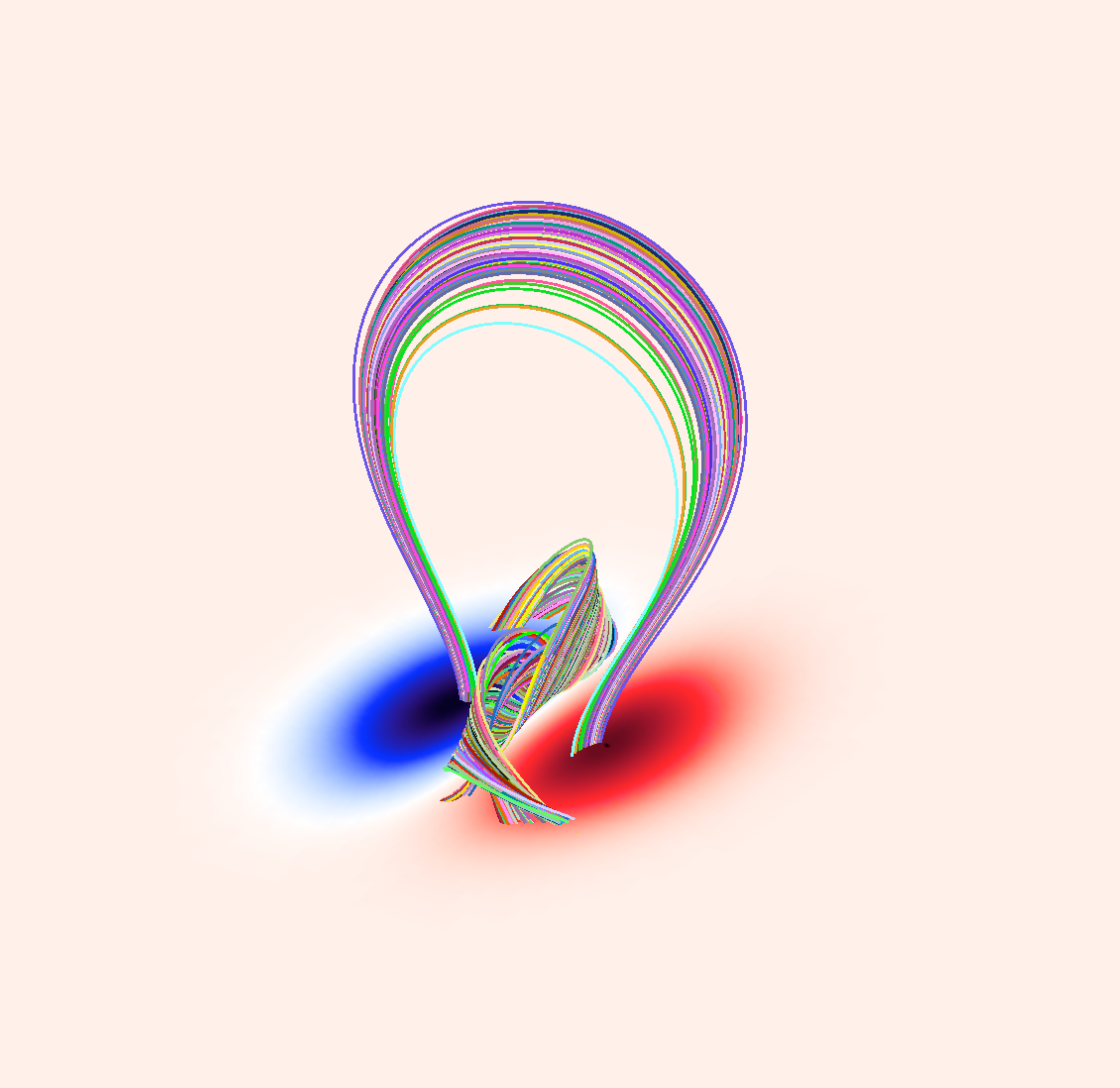}
\caption{Selected field lines of the equilibrium configuration obtained at $t=820$, i.e., 
during the phase of localised flux cancellation.
A TFR exhibiting dips favourable to the support of cool material has formed from the two J-shape 
loops seen on Figure\ref{fig1}.
\label{fig3} }
\end{figure}

(2) During this phase, viscous relaxation always leads to a neighbouring equilibrium. The TFR created along the neutral line grows up both laterally and vertically, as new flux participates to its structure.  As expected for a system in near equilibrium \citep{Aly84,Aly91,Sturrock91}, we find that $\Wpo(t) <W(t) < \Wop(t)$, where $W(t)$ is the magnetic energy of the configuration, and  
$\Wpo(t)$ and $\Wop(t)$ are, respectively, the energies of the potential field $\bBpo(t)$ and the open field $\bBop(t)$ having the same flux distribution on $S_{h}$ as $\B(t)$. 
$W(t)$ and $\Wpo(t)$ are observed to decrease slightly, while the ratio $W(t)/\Wpo(t)$ increases (see Fig. 
\ref{fig2ab}b). This behaviour is expected since $\Wpo(t)$ depends only on the photospheric flux distribution, decreasing by FC, while $W(t)$ depends also of the volumic distribution of coronal currents. As $\Wpo(t)$, $\Wop(t)$ depends only on the photospheric flux distribution and decreases, but the inequality $\Wop(t) > W(t)$
stays satisfied as FC has not yet been very effective.

\begin{figure}
\epsscale{1.}
\plotone{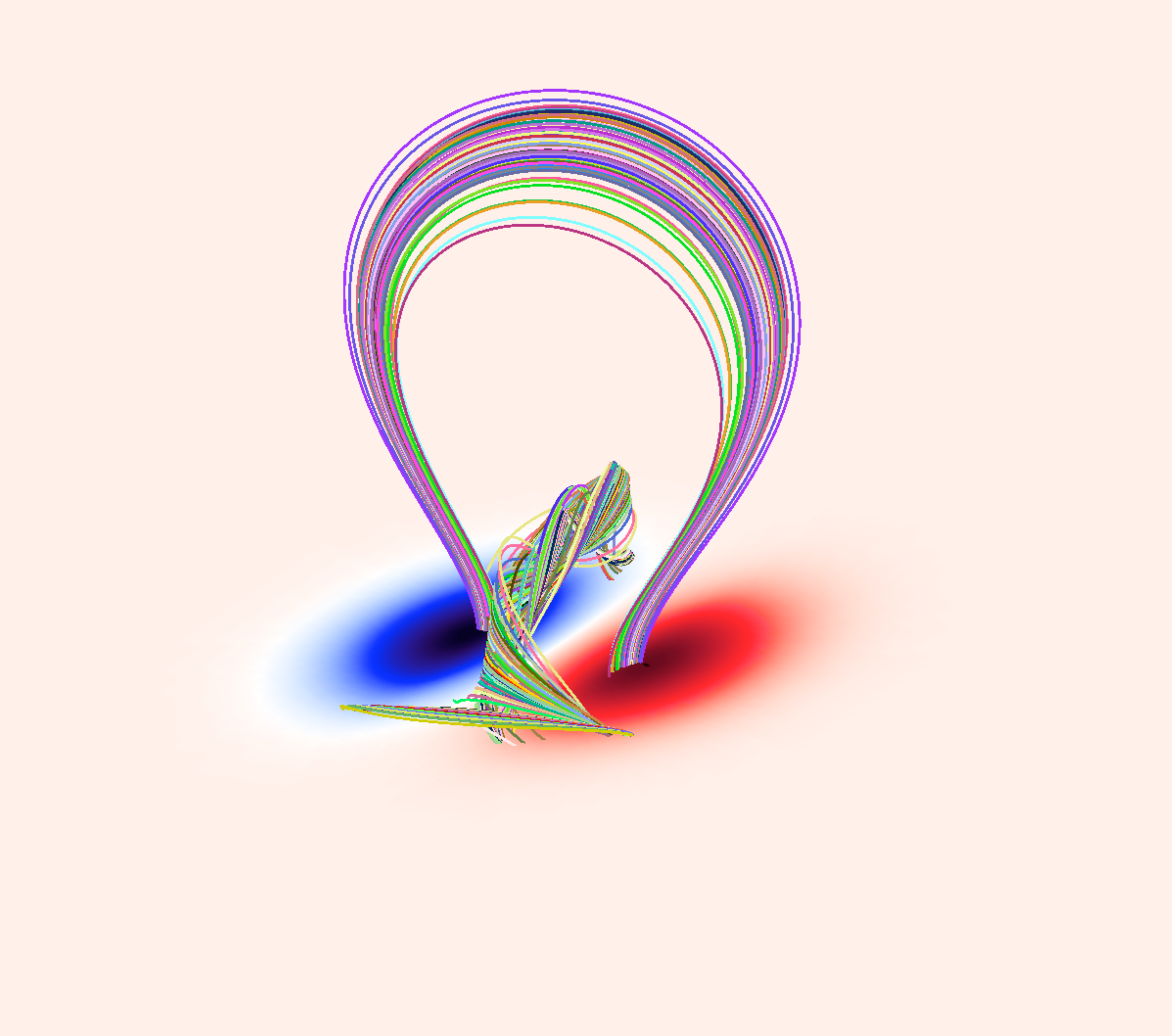}
\caption{Selected field lines of the evolving configuration at $t=838$,
further during the phase of localised flux cancellation .
Beyond this state viscous relaxation to an equilibrium does no longer hold when flux cancellation is switched off.
\label{fig4} }
\end{figure}

\notetoeditor{Figure\ref{fig4} should appear here}

\begin{figure}
\epsscale{1.1}
\plotone{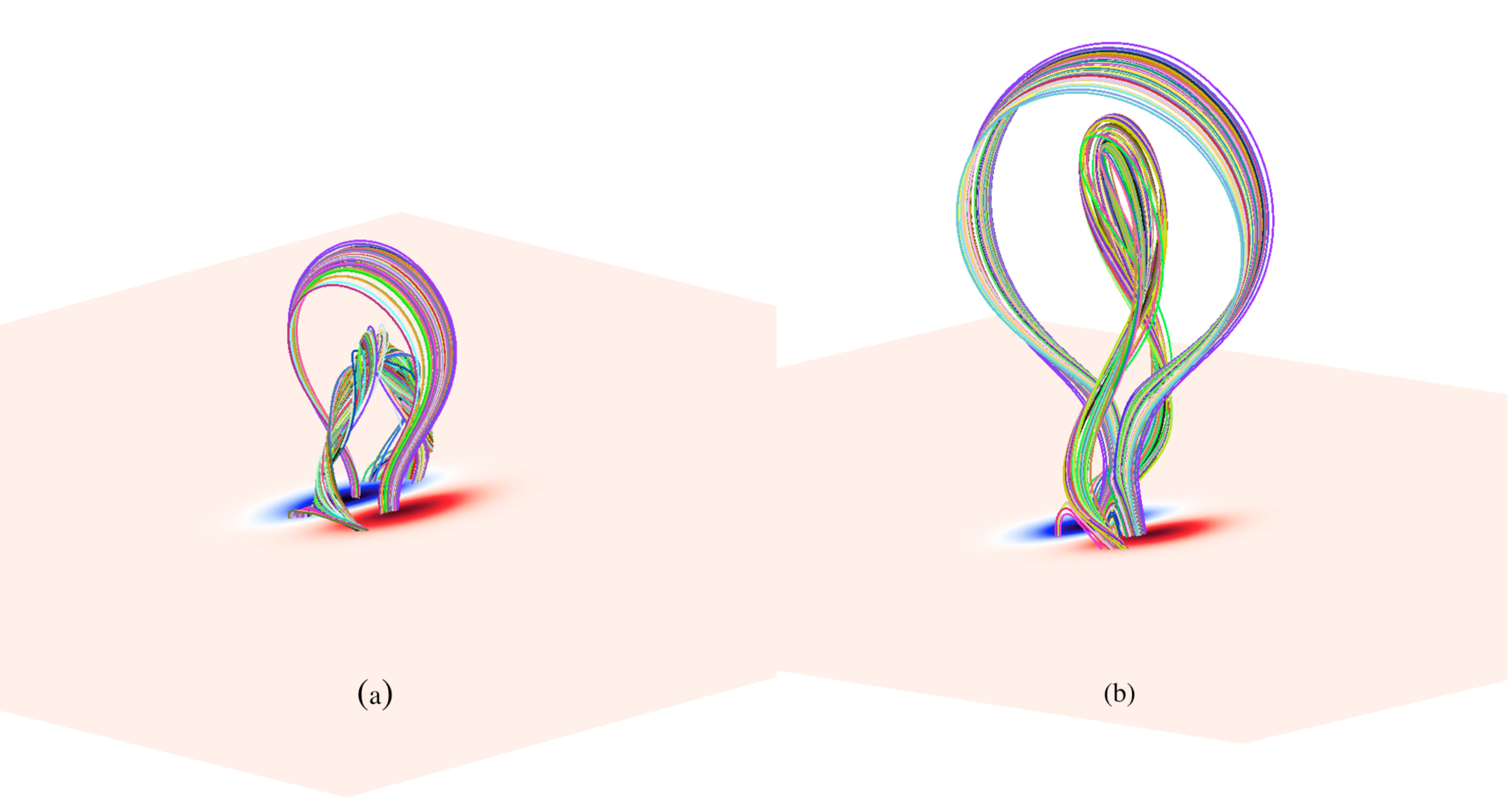}
\caption{Selection of field lines of the configurations obtained at (a) $t=885$ and (b) $t=1135$ 
from the configuration of Figure \ref{fig4}, with FC switched off. 
The global disruption involves opening, reconnection
through the overlying arcade and below, and formation of
a current sheet, associated to a high dissipation of
magnetic energy.
\label{fig5} }
\end{figure}

(3) After a critical time $t_{gd}\approx 838$, the configuration experiences a global disruption, as shown on Figure \ref{fig4}, and FC is switched off. Viscous relaxation does no longer lead to an equilibrium close to $\B(t)$, and the
configuration evolves dynamically towards opening (see Fig.  \ref{fig5}). 
As in \citet{AmariLuAlTa96b,AmariLuMiLi00, AmariLuAlMiLi03a, AmariLuAlMiLi03b,AmariAlMiLi07},  opening is characterized by a transition to very fast expansion suffered by a bundle of lines, which thus close down eventually at very large distances. Along with that process,
reconnection develops through the overlying arcade and a current sheet forms below, associated to dissipation.
\notetoeditor{Figure\ref{fig5} should appear here}
At this stage, since our results rely only on viscous relaxation, it can be either that there exists an accessible equilibrium, but that it is unstable (case UE of our classification) or that such an equilibrium does not exist (case NE of our classification). Discriminating between the two situations could be done in principle by first trying to compute force-free equilibria satisfying the boundary conditions on $B_{n}$ and $\alpha=(\rot\B)_{z}/\Bz$ given by the evolution model before and after $t_{gd}$, respectively (e.g., by using the Grad-Rubin methods developed in \citet{AmariBoAl06}), and by studying their stability when they exist. But this is a difficult task since convergence of the equilibrium algorithms might be difficult to achieve around the critical point, while testing stability requires superimposing perturbations of some types. Then, as we favor a priori NE to be responsible for the disruption, we adopt as in \cite{AmariLuMiLi00, AmariAlMiLi07} 
the strategy consisting to show that a necessary condition for the existence of an equilibrium ceases to be satisfied at $t_{gd}$.
\par

(4) The idea is that it becomes energetically favorable for the system to open once the magnetic energy $W(t)$ starts exceeding the energy of an accessible semi-open field $\B_{SO}(t)$. The $\B_{SO}(t)$ which seems to be relevant here is defined as follows: it has all its lines connected to $supp(SO)$ being open and all the other lines being closed. A priori $supp(SO)$ can be reasonably defined in two different ways: it can be identified either with the part of $S_{h}$ to which the fast expanding lines are connected, or with the part of $S_{h}$ where the twist/shear of the initial force-free field is concentrated. Happily, however, the two definitions select the same region $supp(SO)$. We thus estimate
$W_{SO}(t)$ by equating it to the energy of the potential field $\bBpo'(t)$ satisfying 
$B_{\po z}'=\abs{q_{0}}$ on  $supp(SO)$ and $B_{\po z}'=q_{0}$ on $S_{h}\setminus  supp(SO)$. As guessed, it appears that the transition to NE occurs indeed when 
$W_{SO}(t)$, which decreases faster than $W(t)$, becomes comparable to the latter.  In fact Figure \ref{fig2ab}b shows that, while 
$W(t)/\Wpo(t)$ increases, $W_{SO}(t)/\Wpo(t)$ decreases, and the two curves cross for  
$t\simeq t_{gd}$. From this point, any attempt to find a neighbouring equilibrium fails and the configuration experiences a major disruption as shown on Figure \ref{fig5}.

We note finally that the amount of cancelled flux at $t_{gd}$ is about $6\%$, a small value compared to the $30\%$ found in \cite{AmariLuMiLi00} and even below the $9\%$ found in the quadrupolar case in \cite{AmariAlMiLi07}. 
Note that the initial states considered in these previous works were also quite highly sheared.
We can thus state that a small amount of FC is sufficient to trigger a large disruption. 
But we do not pretend in any case that the $6\%$ value is general. Depending on the initial state, a higher or lower amount of partial cancellation may be needed for a large disruption to be produced. And for low shear initial states, it is even possible that only a confined eruption be obtained as in \citet{AmariLuAlMiLi03b}.

\acknowledgments
TA thanks G. Aulanier for discussion.
We acknowledge support from NASA's
Sun-Earth Connection Theory
Program,  NASA's STEREO/SECCHI Consortium and Centre National
d'Etudes Spatiales. The numerical simulations performed in this paper have
been done on the NEC SX8  supercomputer of the Institute
I.D.R.I.S of the Centre National de la Recherche Scientifique.



\clearpage

\clearpage

\clearpage


\clearpage

\clearpage


\end{document}